# Photoresist-Free Lithography via Light-Induced Redox Patterning of Oxide Thin Films

*Johannes Frantti*, Yukari Fujioka*
Reciprocal Engineering – RE Oy, Jaalaranta 9B 42, Helsinki, Finland
E-mail: johannes.frantti@fre.fi

*Christopher Rouleau, Alexander Puretzky, Ilia N. Ivanov*
Center for Nanophase Materials Sciences, Oak Ridge National Laboratory, Oak Ridge, TN 37831, USA

## Abstract

Lithographic patterning of nanoscale devices typically depends on photoresists, wet processing and multistep pattern-transfer workflows that impose both environmental and scaling constraints. Focused laser irradiation of $(Ni,Co)_{1+2x}Ti_{1-x}O_3$ under a reducing atmosphere drives localized oxygen removal, directly converting the insulating oxide into embedded metallic features without etching or chemical processing. Here we demonstrate a photoresist-free patterning method based on light-induced redox transformations in $(Ni_{0.4}Co_{0.6})_3O_3$ ($x$ = 1) thin films. Our results indicate that the transformation is self-limiting, as the high optical absorption of the emerging metallic phase restricts further energy penetration and confines the reduction front, producing conductive layers with electrical properties comparable to bulk metals. This reversible, orientation-preserving transformation is characteristic of a topotactic redox process, enabling direct nanoscale patterning without photoresists. The process is reversible through re-oxidation and is compatible with standard semiconductor wafer substrates. The role of Ti as a reduction inhibitor is discussed. These results establish a route to direct nanoscale patterning based on controlled redox chemistry, offering a potential alternative to conventional lithography for functional thin-film materials.

## 1. Introduction

Lithographic patterning underpins every modern semiconductor technology, yet it remains tied to workflows built around photoresists, wet chemical processing, and multistep pattern-transfer sequences.[1,2] These processes consume large volumes of ultrapure water and rely on hazardous chemistries and high-global-warming-potential gases such as $SF_6$, $NF_3$, $CF_4$ and $C_2F_6$. Etching steps further require carrier and dilution gases, including helium, a finite and increasingly supply-constrained resource, together with additional fluorinated species to achieve anisotropic profiles. As device dimensions shrink and manufacturing scales expand, the combined dependence on photoresists, toxic etchants, and helium-intensive plasma processes has made the environmental and resource burdens of conventional lithography increasingly difficult to sustain.

Efforts to reduce this complexity have motivated the search for patterning strategies that eliminate photoresists and wet processing altogether. A promising direction is the use of materials that can be reconfigured directly, without masks or etchants, by locally modifying their chemical or electronic structure. However, most candidate systems lack the combination of reversibility, crystallographic stability, and compatibility with semiconductor wafers required for practical device fabrication.

Here we introduce a photoresist-free patterning approach based on light-induced redox transformations in oxide thin films, Figure 1. The method exploits a reversible conversion between an insulating ilmenite-derived oxide[3,4] and a metallic close-packed phase in $(Ni,Co)_{1+2x}Ti_{1-x}O_3$ (NCT)[5]. We use abbreviation NCT to emphasize the presence of Ti.

The reversible oxygen removal and reinsertion while preserving the cation framework is characteristic of a topotactic redox transformation, in the sense originally described by Goodenough for layered and close-packed oxides.[6] The preservation of crystallographic orientation and the reversible ilmenite - metallic close-packed transition observed here are consistent with the topotactic features reported previously for NCT thin films.[5]

The reduction process is initiated through the application of heat to the oxide in the presence of a reducing gas. Previously, we employed resistive heating in conjunction with a mixture of 4%$H_2$+96%Ar to reduce entire films. Here we show that localized laser irradiation in a reducing atmosphere removes oxygen from the lattice, directly writing embedded metallic features without etching, polishing, or water-based cleaning steps. The transformation preserves crystallographic orientation (see XRD pattern in Figure 1 in ref. [5], in which the major reflection from the metallic phase exhibits subsidiary maxima), is reversible through re-oxidation, and is compatible with standard semiconductor wafer materials.

A clear distinction between resistive and laser-driven reduction is rooted in the wavelength-dependent absorption of oxides and metals. The key mechanism we emphasize is the optical response of the metallic phase: once it forms, its strong absorption sharply limits further energy penetration, producing an intrinsically self-confined transformation only a few tens of nanometers deep. This fixes the vertical resolution by material properties, while the lateral resolution is determined solely by the excitation source. We suggest that nanoscale patterning may be achievable using electromagnetic excitation sources such as short-wavelength radiation, where the optically driven self-limiting mechanism remains operative. In addition, highly localized thermal delivery using heated AFM tips can produce sub-20 nm features, offering a complementary route to nanoscale patterning. Taken together, these observations point toward a route for nanoscale lithography based on controlled redox chemistry, without the need for conventional photoresists.

## 2. Material platform

The direct-patterning approach relies on thin-film materials that can reversibly transform between an insulating oxide phase and a metallic close-packed phase. In this work we focus on the ilmenite-derived compound $(Ni,Co)_{1+2x}Ti_{1-x}O_3$ with $x = 1$ and the Ni/Co atomic percentage ratio approximately 40/60 (abbreviated as NCO). Generally, $(Ni,Co)_{1+2x}Ti_{1-x}O_3$ forms over a wide composition range ($-0.25 \leq x \leq 1$) and consists of alternating cation–oxygen layers that accommodate oxygen removal and reinsertion without disrupting long-range crystallographic order.[4,5]
The deposition conditions and structural characteristics of NCT films have been described previously.[3–5] When exposed to a reducing atmosphere, oxygen is selectively removed from the lattice, producing a well-oriented metallic phase. Re-oxidation restores the original insulating structure. NCT films preferentially grow on substrates with hexagonal surface symmetry, which promotes the correct stacking sequence of the ilmenite-derived layers. Examples of substrates that support epitaxial or highly oriented growth are given in ref. [5] and summarized in Table 2 of section 8.2.

## 3. Redox transformation mechanisms

Direct patterning of NCT and NCO films can proceed through two distinct reduction pathways: bulk resistive heating and laser-assisted surface heating. Although both routes ultimately produce a metallic close-packed phase, the underlying physical mechanisms differ significantly.
The redox transformation is accompanied by large changes in functional properties. Under complete film reduction, the *c*-axis undergoes a contraction of about 22%.[5] The insulating oxide exhibits a saturation magnetization of ≈ 100 emu $cm^{-3}$ [4,5] and a resistivity above $10^8 \Omega m$ [4], whereas the metallic Ni–Co phase reaches saturation magnetizations 5–10 times higher and resistivities 16–18 orders of magnitude lower. These contrasts provide a clear physical signature of the phase transformation and enable straightforward electrical and magnetic verification of patterned regions.

### 3.1. Reduction pathways

In both resistive and laser-assisted reduction, metallic-phase formation initiates at the film surface exposed to the reducing atmosphere, consistent with oxygen exchange occurring at the gas–solid interface. This was confirmed using films grown on $LiNbO_3$ substrates, where tensile strain produces a two-layer structure consisting of a strained bottom layer and a relaxed top layer. In our two-layer reduction experiments, we found that the relaxed surface layer vanishes before the strained layer adjacent to the substrate, demonstrating that oxygen removal initiates at the surface and advances inward. Previous work on films grown on sapphire substrates [5] shows that the overall reduction mechanism is similar for resistive and the currently discussed laser heating, but the kinetics differ. Figure 2 compares the optical absorption of NCO and $(Ni_{0.4}Co_{0.6})TiO_3$. Ti-containing films exhibit lower absorption across the visible range, and experimentally the temperature required for resistive reduction increases with increasing Ti fraction.

#### *3.1.1. Reduction in NCO films*

To elucidate the reduction behavior of NCO thin films in hydrogen-containing atmospheres, we draw on prior studies of nickel and cobalt oxides. In ref. [7], NiO reduction proceeds through dissociative adsorption of $H_2$ on Ni sites adjacent to oxygen vacancies, followed by formation and desorption of $H_2O$. The emergence of small Ni clusters facilitates further reduction. Cobalt follows a similar pathway: $Co_3O_4$ first reduces to CoO, and only then to metallic cobalt, initially in the hexagonal close-packed form and, above ~400 °C, in the face-centered cubic structure.[8]
The essential distinction for NCO is structural. The Ni–Co cation framework in the ilmenite-derived oxide already lies within small displacements of a hexagonal close-packed arrangement (Figure 1). As a result, oxygen removal directly produces an hcp-type metallic lattice without requiring intermediate structural rearrangements. This direct oxide-to-hcp transition is not possible in NiO, whose rocksalt structure bears no relation to a close-packed metal, and even $Co_3O_4$ must pass through CoO before reaching metallic cobalt. In contrast, NCO's pre-existing close-packed cation skeleton enables a direct and reversible NCO → hcp-(Ni,Co) transformation, which is the central functional property exploited in the present direct-patterning approach.

#### *3.1.2. Influence of titanium on reduction temperature*

The NCT structure is derived from the ilmenite and corundum lattices, both of which contain three oxygen atoms per two cations. In the fully substituted case (x = 1), the lattice contains three transition-metal cations (Ni and Co) per three oxygen atoms, meaning that all octahedral sites are occupied.[3] Alloying with Ti provides a tunable parameter for adjusting physical properties and, importantly, offers a means to increase the reduction temperature. Consequently, Ti-rich layers can act as barriers that suppress laser-assisted reduction in the underlying material. Increasing the $Ti^{4+}$ fraction lowers the total number of reducible transition-metal cations and alters the local oxygen coordination environment. We suggest the reduction temperature increases through four distinct mechanisms.
Stronger Ti–O bonds and higher vacancy formation energy
$Ti^{4+}$ forms stronger Ti–O bonds than Ni/Co in their reducible valence states. Consequently, the energy cost to remove an oxygen is higher when Ti is present. A higher oxygen-vacancy formation energy directly raises the thermal activation barrier for reduction and therefore the temperature required for a given reduction rate.

*Lower fraction of reducible cations*

Substituting Ti for Ni/Co decreases the number of reducible transition-metal cations per oxygen, since $Ti^{4+}$ does not readily reduce to a metallic state under the same conditions. With fewer reducible cations available, more oxygen must be removed per reducible cation to reach the same metallic fraction. This increases the net thermal work required and thus elevates the reduction temperature.

*Reduced electronic conductivity and charge carrier density*

Charge carriers play a central role in charge compensation during reduction. Ni/Co oxides that reduce readily typically exhibit enhanced small-polaron or electronic conductivity once oxygen vacancies form. Ti-rich compositions remain more insulating because $Ti^{4+}$ does not contribute itinerant carriers.[4] Lower electronic conductivity slows charge compensation and oxygen migration, requiring higher temperatures to reach comparable vacancy concentrations.

*Lower optical absorption and reduced volumetric heating under resistive conditions*

Absorption data in Figure 2 show that Ti-containing films absorb less light across the spectrum. Under resistive heating this matters indirectly: lower optical absorption is not the primary factor for resistive heating, but it correlates with the electronic energy band structure changes (there are fewer free carriers due to the larger band gap) that also make thermal reduction harder. Under laser heating the situation differs because local photothermal and photochemical pathways can concentrate energy at the surface and drive reduction even when the temperature of the bulk is below the thermal activation temperature.

*Note on Ti-content upper limit*

Our tests show that NCT films with $x \approx 0.5$ still reduce to a close-packed metallic phase. However, introducing $Ti^{4+}$ replaces two divalent cations, creating vacant octahedral sites.[3] At sufficiently high Ti concentrations, these vacancies disrupt the continuity of the close-packed cation framework. Consequently, a fully connected close-packed metallic structure cannot be preserved once the Ti content becomes large, and the reduced phase will differ structurally from the metallic phase formed in Ti-free NCO films.

*3.1.3. Laser-assisted reduction*

Laser-assisted reduction is governed by surface-localized photothermal and photochemical processes. Figure 3 shows Raman spectra from unprocessed, slightly reduced, and fully reduced regions of an NCO film on *z*-cut $LiNbO_3$. In the unprocessed and slightly reduced regions, Raman peaks from the substrate dominate, whereas in the fully reduced region the substrate signal is absent. This indicates that the metallic layer is sufficiently thick and optically absorbing to prevent the probing laser from reaching the underlying oxide.
The spatial heat distribution along the film thickness during laser irradiation is approximated as exponentially decaying with depth. To give a quantitative number, we applied Beer-Lambert's law to the NCO film (thickness 90nm) at 405nm, where the measured absorbance is 80%, yields an absorption coefficient $\alpha = 1.8\times 10^7 m^{-1}$, corresponding to an absorption length $l_{abs} = 1/\alpha \approx 56$nm. While high oxide absorption is necessary to initiate surface heating, the final thickness of the metallic layer is governed by the absorption properties of the metal. As a first approximation, the metallic-layer thickness can be taken as the optical absorption length of the Ni–Co alloy. Table 1 lists the refractive index *n*, extinction coefficient *k*, and absorption coefficient $\alpha$ for Ni and Co, extracted from ref. [9]. The composition-averaged absorption length is approximately 11 nm. This value likely represents a lower bound, since thermal diffusion and continued oxygen removal can extend the reduction front beyond the optical penetration depth. Below, we provide an independent thickness estimate based on the XRD peak broadening.

The lack of substrate signal in the spectrum measured from the metallic surface, shown in Figure 3(a), is consistent with this: the absorption length at 532 nm is slightly less than at 405 nm, and thus the laser light cannot penetrate to the oxide layer.

Table 1. Refractive indices, extinction coefficient and absorption coefficient $\alpha = \frac{4\pi}{\lambda}$ of Ni and Co at $\lambda$ = 405nm and 532 nm.[9]

| | Ni | Co |
|---|---|---|
| $n$ (405 nm) | 1.7100 | 1.5900 |
| $n$ (532 nm) | 1.8775 | 2.0014 |
| $k$ (405 nm) | 2.6300 | 2.9900 |
| $k$ (532 nm) | 3.4946 | 3.7350 |
| $\alpha$ ($10^7$ $m^{-1}$) (405 nm) | 8.16 | 9.28 |
| $\alpha$ ($10^7$ $m^{-1}$) (532 nm) | 10.8 | 11.6 |

*Local versus bulk reduction pathways*

Resistive heating produces a bulk thermal reduction pathway, in which oxygen-vacancy formation and diffusion occur throughout the film thickness. The higher vacancy-formation energy, lower fraction of reducible cations, and poorer electronic/ionic transport in Ti-containing films all increase the temperature required to reach a given vacancy concentration and metal fraction.

Laser-assisted reduction, in contrast, is a highly localized, non-equilibrium pathway driven by surface-confined photothermal excitation. Energy deposition is concentrated within the top few nanometers, allowing the reduction front to advance even when the underlying film remains below the bulk thermal activation threshold. This surface-localized mechanism can partially circumvent the thermodynamic penalties introduced by Ti, enabling reduction at lower effective temperatures and producing metallic layers whose thickness is governed by the optical absorption of the emerging metal. The suggested self-limiting nature of the reduction process has important implications for patterning resolution. As the metallic phase forms, its high optical absorption suppresses further energy penetration into the film, restricting the transformation to a shallow surface region of approximately 10–20 nm, as estimated from the extinction coefficient. This intrinsic depth confinement is largely independent of the lateral dimensions of the excitation source. Consequently, the achievable lateral resolution is further suggested to be governed not by thermal diffusion within the film but by the spatial extent of the energy input. In combination with ultrathin films and highly localized excitation sources, such as heated atomic force microscopy (AFM) probes, this mechanism provides a potential route toward nanoscale patterning. Existing scanning-probe lithography techniques have demonstrated feature sizes below 20 nm, indicating that the present redox-based approach could be extended to similar or smaller dimensions. Moreover, the use of shorter-wavelength radiation, such as extreme ultraviolet, may enable parallel patterning at comparable length scales.

**4. Patterning examples**

Figure 4 shows metallic patterns formed by laser-assisted reduction of a NCO film grown on a *z*-cut $LiNbO_3$ substrate. The reduced regions appear as embedded metallic features within the oxide matrix. An XRD scan collected from the patterned area [Figure 4(c)] shows a distinct and nearly symmetric reflection at 44.4°, characteristic of the close-packed Ni–Co metallic phase (compare with Figure 5), while the dominant peaks originate from the insulating oxide and the substrate. The symmetry of this metallic reflection indicates that strain does not significantly distort the peak profile.

To evaluate the electrical properties of the patterned structures, we wrote parallel metallic lines, each 5 µm wide and 1 mm long. The first three lines were spaced 50 µm apart, the fourth line was placed 100 µm from the third, and the fifth line was positioned 50 µm beyond the fourth. The metallic layer thickness, estimated from the Scherrer analysis of the XRD peak broadening, is approximately 20 nm. We note that Scherrer analysis conflates size broadening with strain broadening. In fully reduced films produced by resistive heating, the diffraction peak profile is symmetrical (see Fig. 2 in ref. [5]), indicating minimal strain-induced distortion. In the present case, however, the metallic stripe is embedded within an oxide matrix, and the strain state of the metallic regions may differ from that of a uniformly reduced film. This strain–size convolution is expected to broaden the diffraction peak and can therefore lead to an underestimation of the metallic-layer thickness. Electron-beam-deposited contact pads were added to ensure reliable electrical connection to the six lines in parallel.

The measured resistance of 54 Ω corresponds, for the given geometry, to a resistivity of $2.8 \times 10^{-8}$ Ω·m and thus a conductivity of $3.6 \times 10^{7}$ S·m$^{-1}$. This value exceeds the commonly reported room-temperature conductivities of bulk nickel ($1.4 \times 10^{7}$ S·m$^{-1}$) and cobalt ($1.6 \times 10^{7}$ S·m$^{-1}$) [10–12]. The minimum conductivity can be estimated by assuming that the metallic layer thickness equals the full film thickness. Using the independently measured thickness of similarly grown films (74 nm) yields a lower-bound conductivity of $9.8 \times 10^{6}$ S·m$^{-1}$. Taken together, these results confirm that the patterned regions exhibit conductivities comparable to, and potentially exceeding, those of bulk nickel and cobalt. Further work is underway to refine the thickness estimates and fully quantify this behavior. The measured resistance of 54 Ω verifies the formation of a metallic phase, and the conductivity is consistent with transport in the plane perpendicular to the hexagonal *c*-axis (or cubic ⟨111⟩ direction), where reduced grain-boundary scattering and enhanced carrier mobility are expected.
The resistance of the unprocessed thin film was above the present measurement scale. Typical thin film resistance values, measured for a system dedicated for high-resistances, are composition dependent and were found to vary between $2 \times 10^{8}$ to $1.70\times 10^{10}$ Ω·m.[4]

**5. Re-oxidation**

The metallic phase produced by reduction can be re-oxidized to recover the original insulating oxide structure, see Figure 5. To demonstrate reversibility, we re-oxidized the reduced NCO film described in ref. 5. The sample was first reduced in a 4% $H_2$ + 96%Ar atmosphere at 574 °C for 28 min. Figure 5 shows the characteristic metallic Ni–Co reflection at 44.38°, confirming the formation of the close-packed metallic phase. The sample was subsequently re-oxidized in the deposition chamber at 10 mTorr $O_2$ and 574°C for 32 min. The XRD pattern collected after re-oxidation reveals the reappearance of the insulating oxide reflections, demonstrating successful recovery of the oriented oxide phase (hexagonal *c*-axis perpendicular to the substrate plane), with no indications of secondary orientations or residual metallic reflections.
For reference, an XRD pattern measured from a film grown under identical conditions to the sample used in the redox experiments is also shown in Figure 5. Apart from the (006) reflections from the film and substrate, no additional peaks are observed.
These results confirm that the reduction process is reversible and that the insulating oxide structure can be restored under appropriate oxidation conditions.

**6. Future prospects**

Future work will focus on quantifying the depth profile of the metallic conduits by transmission electron microscopy, evaluating thin-film device performance, and upgrading the optical system and light source to achieve narrower linewidths. The patterning process will also be refined to enable controlled re-oxidation of selected regions directly under the laser beam. As illustrated in Figure 6, embedded planar conductors of virtually any shape can be produced within the electrically insulating thin-film host. Beyond passive RLC components, the same approach can be extended to realize metasurfaces and waveguides, including spin-wave structures, and - through multilayer integration - more advanced device architectures. For example, oriented NCO thin films grow epitaxially on the semiconducting phase of $V_2O_3$, providing a promising platform for multilayer implementations such as field-effect-based devices.

Reduction experiments on titanium-alloyed samples, conducted in both single-layer and multilayer configurations, confirm the formation of a close-packed Ni–Co–Ti metallic phase similar to that observed in NCO films, albeit at higher temperatures. Laser-based patterning of Ti-alloyed thin films has likewise been successfully demonstrated, and further work is underway to quantify the process. The role of lattice mismatch between the film and the substrate influences the strain state of the oxide film and affects the crystalline properties of the resulting metallic layer. We have reduced entire NCO and NCT films grown on various substrates at different temperatures and find that the substrate and reduction temperature together determine whether a hexagonal close-packed or cubic close-packed metallic structure forms. This structural outcome, in turn, affects the re-oxidation behavior of the films. In general, lower-temperature processing favors improved reversibility, implying that the characteristics of local laser-induced reduction differ from those of full-film reduction.

## 7. Conclusions

We demonstrate a direct-writing protocol that produces embedded metallic patterns within NCT thin films through spatially selective laser-induced reduction. The reversible oxide–metal transformation enables pattern formation without photoresists, wet etching or electromechanical polishing, and is compatible with standard semiconductor wafer materials. This approach provides a general platform for generating functional metallic features in magnetic, semiconducting and electro-optic thin-film systems. Although device-level implementations lie beyond the scope of this study, the results establish a versatile and sustainable route for patterning functional oxide films and point toward redox-based nanoscale lithography

## 8. Methods

### 8.1. Direct patterning instrument

A direct-write platform enabling deterministic formation of embedded metallic, magnetic, insulating, and optical structures within continuous thin films was designed, constructed, and validated. The patterning system uses a 405 nm laser fiber-coupled to a UV objective through a single-mode, polarization-maintaining fiber. Beam-delivery optics allow simultaneous illumination and camera-based viewing of the sample. The resulting laser spot size on the film is approximately one micron, set by the fiber and objective configuration.

### 8.2. Thin-film deposition and substrate requirements

Device fabrication proceeds in two steps. First, insulating $(Ni_{0.4}Co_{0.6})_3O_3$ (NCO) films were deposited by pulsed laser deposition as described in refs. [3–5]. We have grown and characterized between 500 and 1000 thin-film samples spanning a wide range of compositions, thicknesses, and substrates. This extensive dataset allows us to tune the deposition parameters to reliably produce films with the desired structural and functional properties. Typical film thicknesses were 70 - 100 nm depending on substrate. The subsequent step in the process is the patterning step, which is described in Section 8.3. Films were grown on a range of substrates to evaluate the role of substrate orientation, lattice symmetry, and strain state in supporting the ilmenite-derived stacking sequence. In all cases, the films adopt *c*-axis orientation perpendicular to the substrate plane, consistent with the close-packed oxygen-layer stacking required for the topotactic transformation. Substrates with hexagonal or pseudo-hexagonal surface symmetry (e.g., sapphire, $LiNbO_3$, SiC, GaN, AlN, α-quartz) reliably support oriented growth, whereas cubic substrates such as Si(111) provide a compatible (111) surface that mimics hexagonal symmetry. The strain state varies with substrate lattice parameters: $LiNbO_3$ induces tensile in-plane strain, GaN/AlN induce compressive strain, and sapphire, Si(111), and α-quartz yield nearly strain-free films. Ti-alloyed NCT films exhibit the same orientation relationships, though the symmetry is lowered due to Ti substitution. These observations indicate that surface symmetry, oxygen-layer stacking compatibility, and moderate lattice mismatch are the key parameters governing successful epitaxial or highly oriented growth.

Table 2. Substrate orientation, strain state, and film characteristics for NCO and NCT films. Strain was evaluated when distinguishable layers were present, consisting of a thin (~15 nm) strained layer and a thicker relaxed layer.

| Substrate | Orientation | Film Orientation | Strain State | Notes |
|---|---|---|---|---|
| $Al_2O_3$ | $(11\bar{2}0)$ | Typically *c*-axis ⊥ substrate. In a film grown at low temperature, two orientations revealed by the $(11\bar{2}0)$ and (0001) reflections. | No strain | Single layer structure (no strained/relaxed bilayer) |
| $Al_2O_3$ | $(10\bar{1}0)$ | *c*-axis ⊥ substrate | No strain | Same characteristics as above |
| $Al_2O_3$ | (0001) | *c*-axis ⊥ substrate | No strain | Same characteristics as above |
| $LiNbO_3$ | *x* cut | Two orientations, revealed by (0001) and $(10\bar{1}1)$ reflections | Tensile in-plane strain | Tensile strained and relaxed (0001)-oriented layers, *c* axes ⊥ substrate. |
| $LiTaO_3$ | *x* cut | Two orientations, revealed by (0001) and $(10\bar{1}1)$ reflections | No strain observed | - |
| $LiNbO_3$ | *y* cut | Two orientations, revealed by a strong $(10\bar{1}1)$ reflection and a very weak $(10\bar{1}1)$ reflection | No strain observed | - |
| $LiTaO_3$ | *y* cut | Two orientations, revealed by a strong $(10\bar{1}1)$ reflection and a very weak (0001) reflection | No strain observed | - |
| $LiNbO_3$ / $LiTaO_3$ | *z* cut | *c*-axis ⊥ substrate | Tensile in-plane strain | Strained and relaxed layer present. *a*,*b* axes tensile, *c* axis compressive |

| Si | (111) | *c*-axis ⊥ substrate | No strain | Film is (0001) oriented; compatible with hexagonal stacking |
|---|---|---|---|---|
| SiC 4H | (0001) | *c*-axis ⊥ substrate | Slight tensile strain in samples grown at 465 °C | Ilmenite-type (0003*n*), *n* odd integer, reflections at low temperature. No strain in samples grown at 595 °C |
| GaN | (0001) | *c*-axis ⊥ substrate | Compressive in plane strain | Strained and relaxed layer present. *a*,*b* axes compressive, *c*-axis tensile |
| AlN | (0001) | *c*-axis ⊥ substrate | Compressive in plane strain | Same strain characteristics as GaN |
| $\alpha$-$SiO_2$ | (0001) | *c*-axis ⊥ substrate | No strain | - |

### 8.3. Device fabrication by laser-assisted reduction

For patterning, the sample is placed on an XY-stage inside a chamber with a controlled reducing atmosphere. The 405 nm laser is delivered through the fiber-coupled objective, and its power is adjusted to initiate local reduction. Pattern formation is verified optically and electrically. Processing is carried out in a 4% $H_2$ + 96% Ar atmosphere, supplied either premixed or from separate gas lines. While 4% $H_2$ reliably initiates reduction, higher $H_2$ fractions were occasionally used. To maintain a stable local reducing environment, the sample is enclosed in a container leaving a narrow gap for gas outflow. The inflow is adjusted to match this gap so that the reducing atmosphere remains constant at the laser-irradiated spot. Because the effective reduction conditions depend on the local $H_2$ concentration rather than the nominal flow rate, the gas inflow is tuned accordingly. All processing is performed at room temperature.

### 8.4. Measurements

Raman measurements were performed in a custom-built setup under a microscope (Olympus) in a backscattering configuration. A 532 nm laser (Cobolt, Hubner Photonics) light was focused onto the sample with a 100x microscope objective (numeric aperture N/A = 0.9) to a spot size of about 1 μm. The excitation laser power on the sample was ~ 0.5mW. The scattered Raman light was analyzed by a spectrometer (Spectra Pro 2300i, Acton, f = 0.3 m, equipped with 1800 grooves/mm grating) that was coupled to a microscope and equipped with an 1800 groves/mm grating and a CCD camera (Pixis 256BR, Princeton Instruments).

Film phase identification and thickness estimations were based on XRD measurements by PANalytical X'Pert Pro MRD instrument equipped with 1.8kW Cu Kα source.

Electronic absorption spectra were acquired between 220 and 800 nm using Cary5000 spectrophotometer (Agilent) in two beam configurations, referencing measurement to a zero absorption (fully blocked sample path and zero absorption (no sample in the beam path). Electronic absorption spectra of substrate and substrate with deposited films were measured versus open reference beam path. A series of measurements were conducted in the ultraviolet-visible range for thin films that had been deposited on substrates of two-side polished *c*-plane sapphire. In addition, a reference measurement was conducted on an identical substrate.

Electrical characterization was performed using a four-point configuration in which the current and voltage leads ($I^+/V^+$ and $I^-/V^-$) were placed on opposite contact pads. Contact quality was verified by measuring the resistance between the $I^+$–$V^+$ and $I^-$–$V^-$ pairs prior to the four-point measurement. Once good contact was confirmed, the measured resistance between the two pads corresponded to the parallel resistance of the five metallic lines, and the resistance of a single line was obtained by multiplying the measured value by five. The deposition of contact pads on the patterned film was facilitated by means of a shadow mask.

**Acknowledgements**

This work was supported by Reciprocal Engineering - RE Oy. Experiments were conducted at the Center for Nanophase Materials Sciences, a DOE Office of Science User Facility at Oak Ridge National Laboratory.

**Conflict of Interest**

The authors declare no competing financial or non-financial interests.

**Data Availability**

The data supporting the findings of this study are available from the corresponding author upon reasonable request.

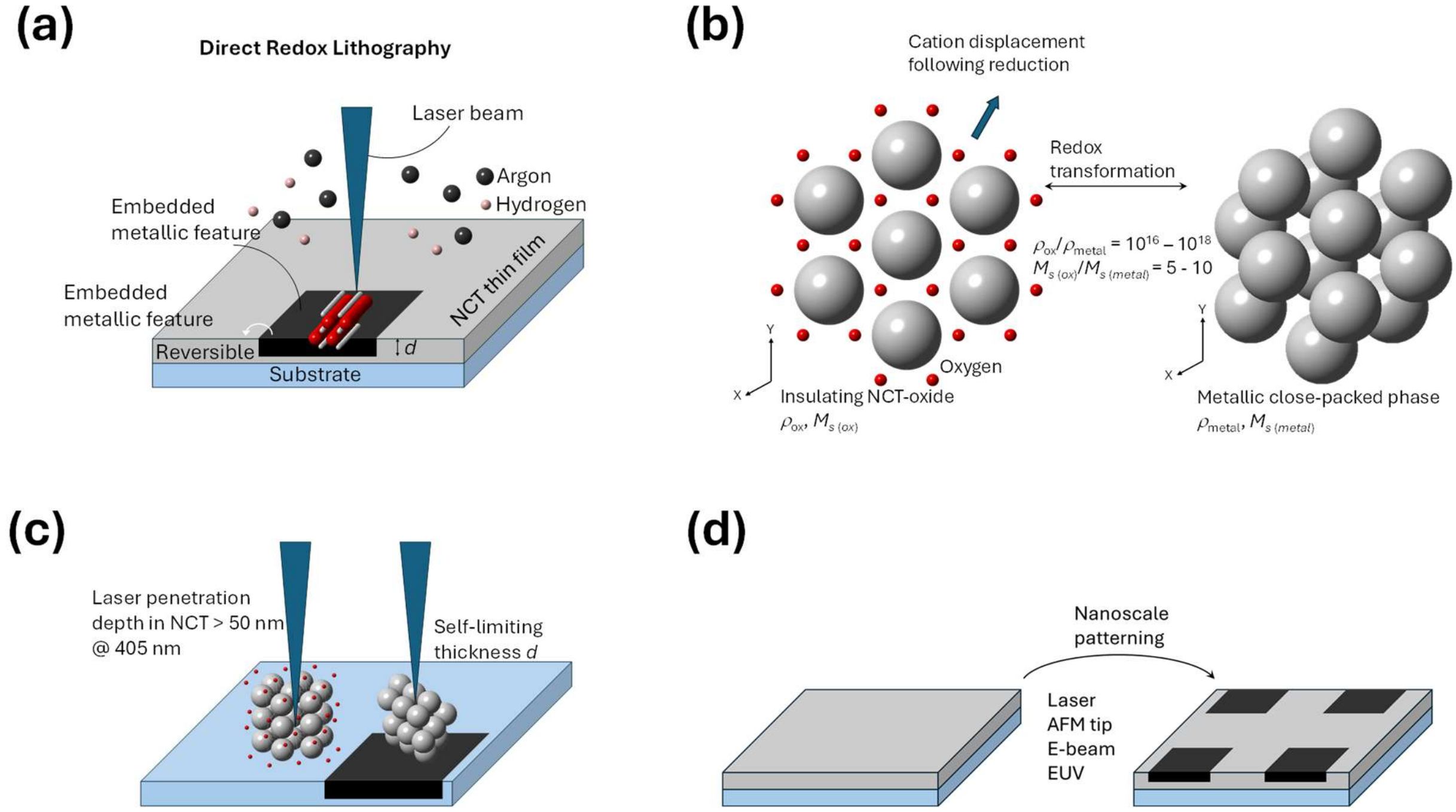

Figure 1. Direct redox lithography enables embedded nanoscale metallic patterning in insulating oxide films. (a) A focused laser beam in a reducing atmosphere ($H_2$+Ar) locally transforms the insulating NCT thin film into an embedded metallic feature. The process is reversible via re-oxidation. (b) Atomic-scale mechanism: oxygen removal and cation displacement convert the ilmenite-derived oxide into a metallic close-packed phase while preserving in-plane orientation. (c) The transformation is self-limiting in-depth *d* (of the order of 10 nm) due to strong optical absorption of the metallic phase, despite deep laser penetration in the oxide. (d) The method is compatible with nanoscale patterning using various excitation sources (laser, AFM tip, e-beam, EUV), with vertical resolution governed by redox physics and lateral resolution determined by the excitation tool.

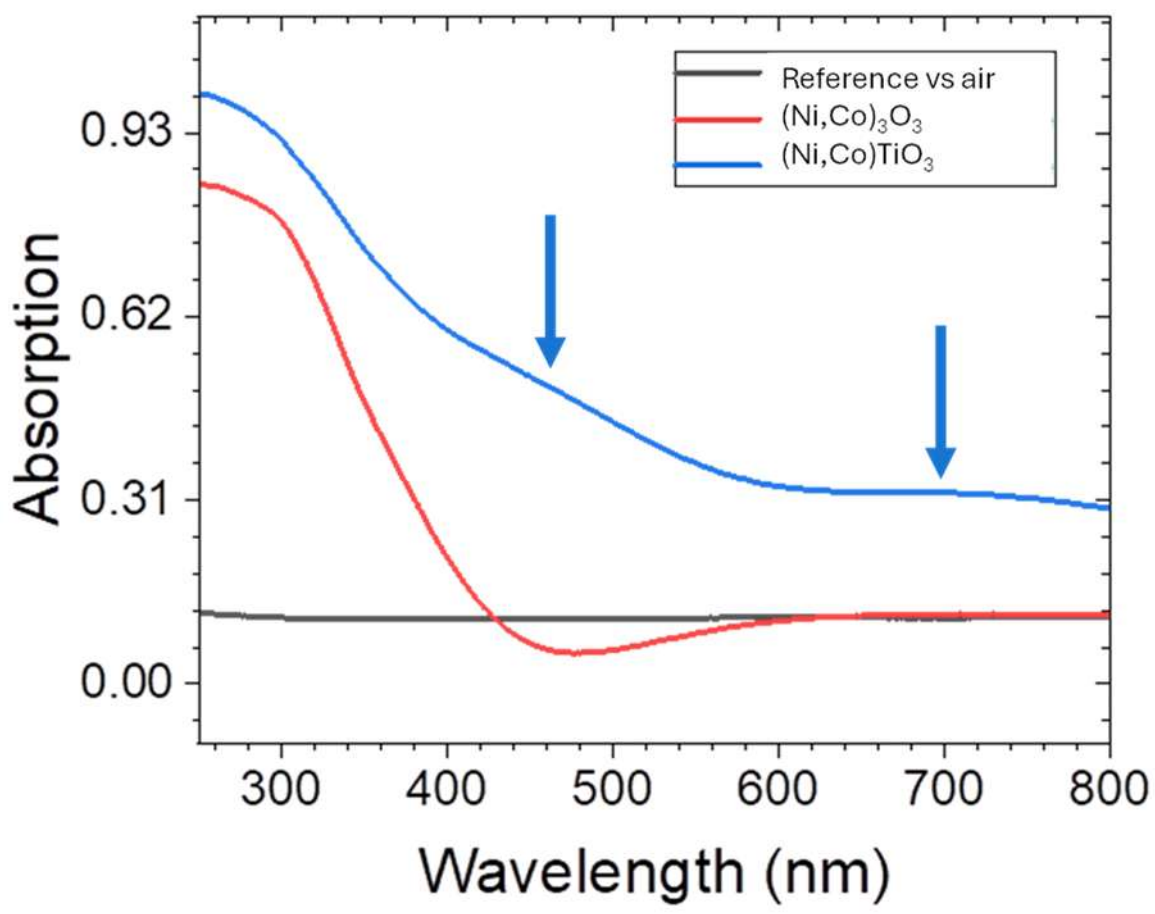

Figure 2. UV-Vis absorption data collected on NCT thin films. Larger light absorption was observed for $(Ni_{0.4}Co_{0.6})_3O_3$ film (NCO, thickness 90nm) (blue line) with broad peak around 300nm and two shoulder around 450 and 700nm, indicated by arrows. Red line shows the absorption in $(Ni_{0.4}Co_{0.6})TiO_3$ film (thickness 63nm) is smaller, with similar broad feature around 300nm, but lacking shoulder peaks. The nearly horizontal line is the reference line, which was measured from the clean double polished $Al_2O_3$ (0001) substrate. The substrate shows insignificant, wavelength independent light absorption between 220 and 800nm.

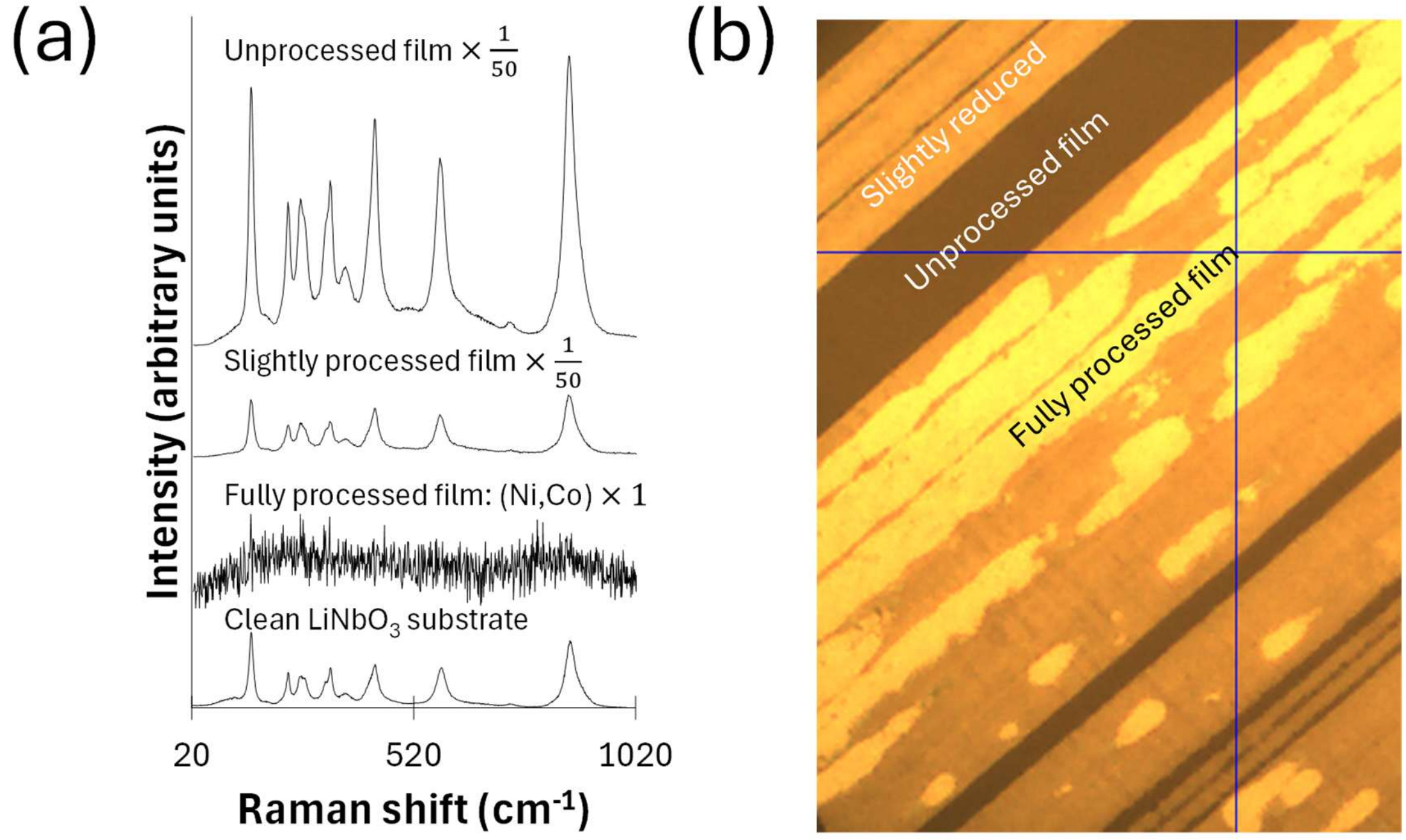

Figure 3. Panel (a) displays Raman spectra excited by 532nm laser from the unprocessed [brown color in optical image in (b)], slightly processed [light brown in image (b)] and metallic [fully processed, yellow areas in image (b)] areas. The intensity values of the spectra collected on unprocessed and slightly processed areas were divided by 50. The intensity from the substrate was practically absent from the spectra measured from fully processed areas. The intensity measured from the unprocessed area was about 20% from the intensity measured from the unprocessed area.

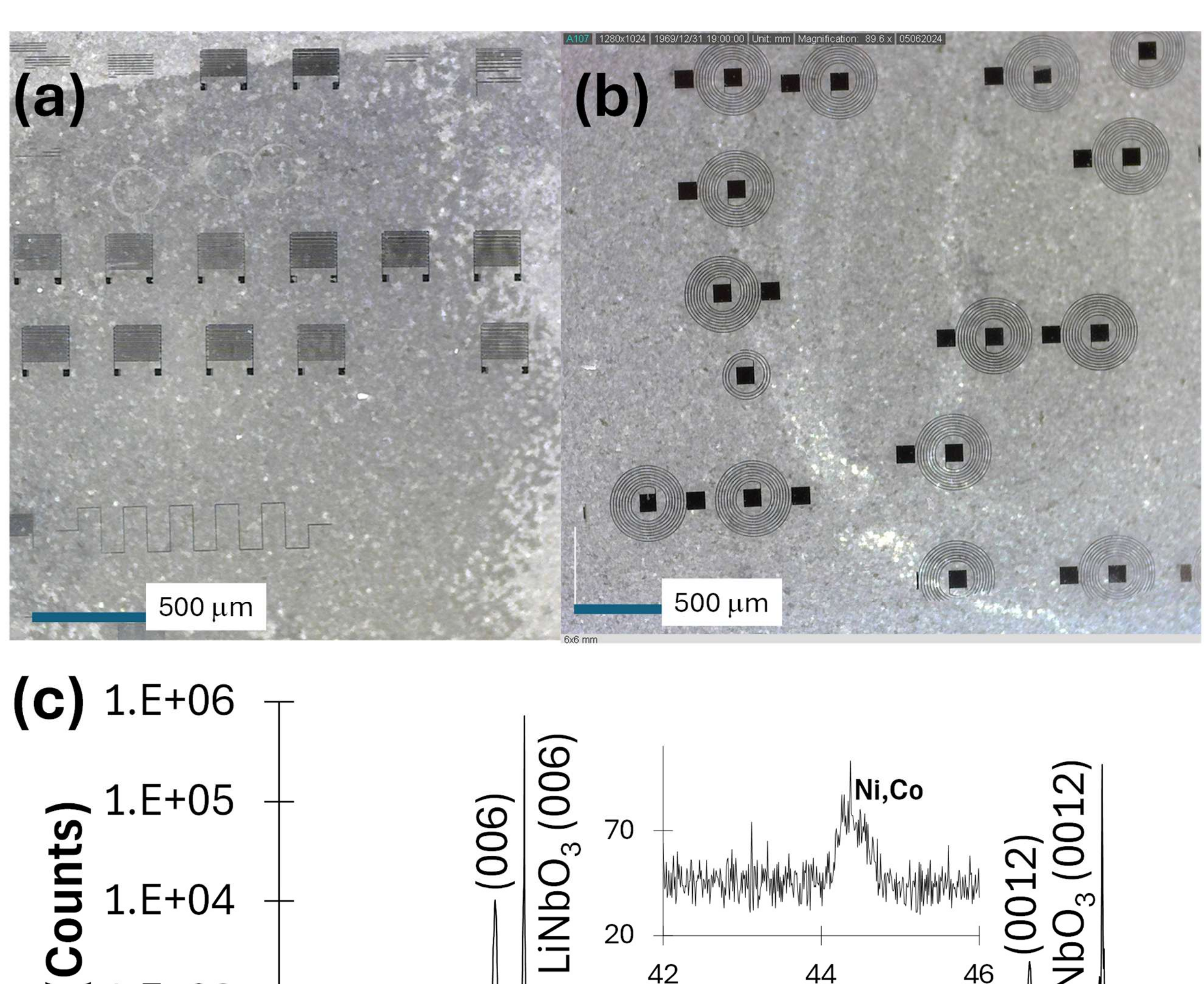

Figure 4. Panels (a) and (b) show optical microscope images of the test patterns created by 405nm laser under reducing gas atmosphere. Panel (c) shows XRD patterns collected on the sample shown in panel (a). The metallic phase revealed itself as a peak at 44.4° (inset). The dominant reflections are from the insulating NCO phase and the substrate. The film thicknesses of both samples, grown under identical conditions, is estimated to be 82nm. Estimate is based on the film thickness grown under the same growth conditions on a similar substrate.

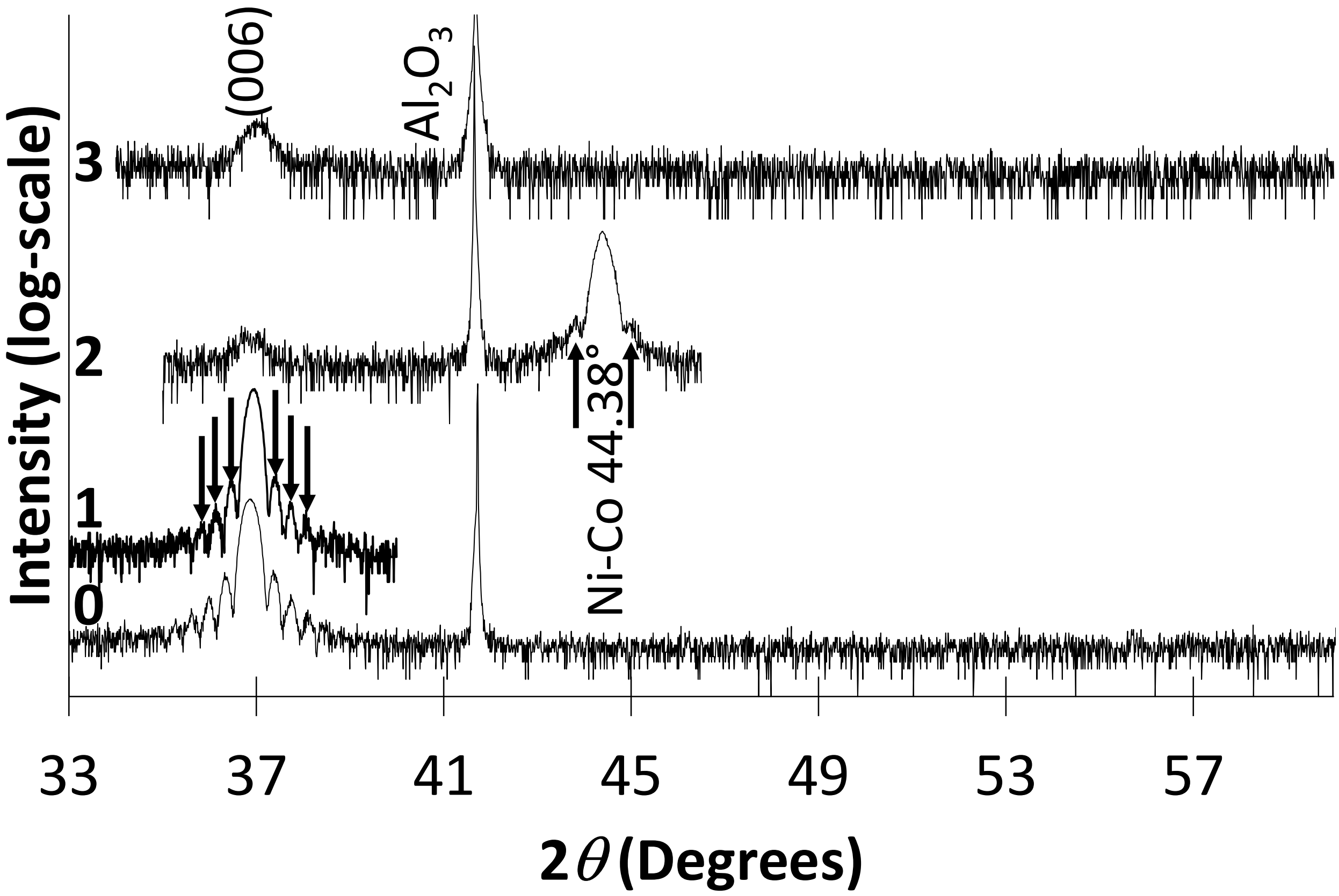

Figure 5. XRD patterns demonstrating the reversibility of the reduction process in NCO films grown on $Al_2O_3$ (0001). Pattern 1 corresponds to the as-grown sample (thickness 22 nm[5]), pattern 2 to the reduced film (thickness 17nm[5]), and pattern 4 to the re-oxidized film. Further details on this sample are provided in Ref. 5. The reduced film exhibits the metallic Ni–Co close-packed phase peak at 44.38°, whereas re-oxidation restores the insulating oxide phase. Arrows mark subsidiary maxima. Pattern 0 is shown for reference and was measured from a film grown under identical conditions to those used for the sample in the redox experiments.

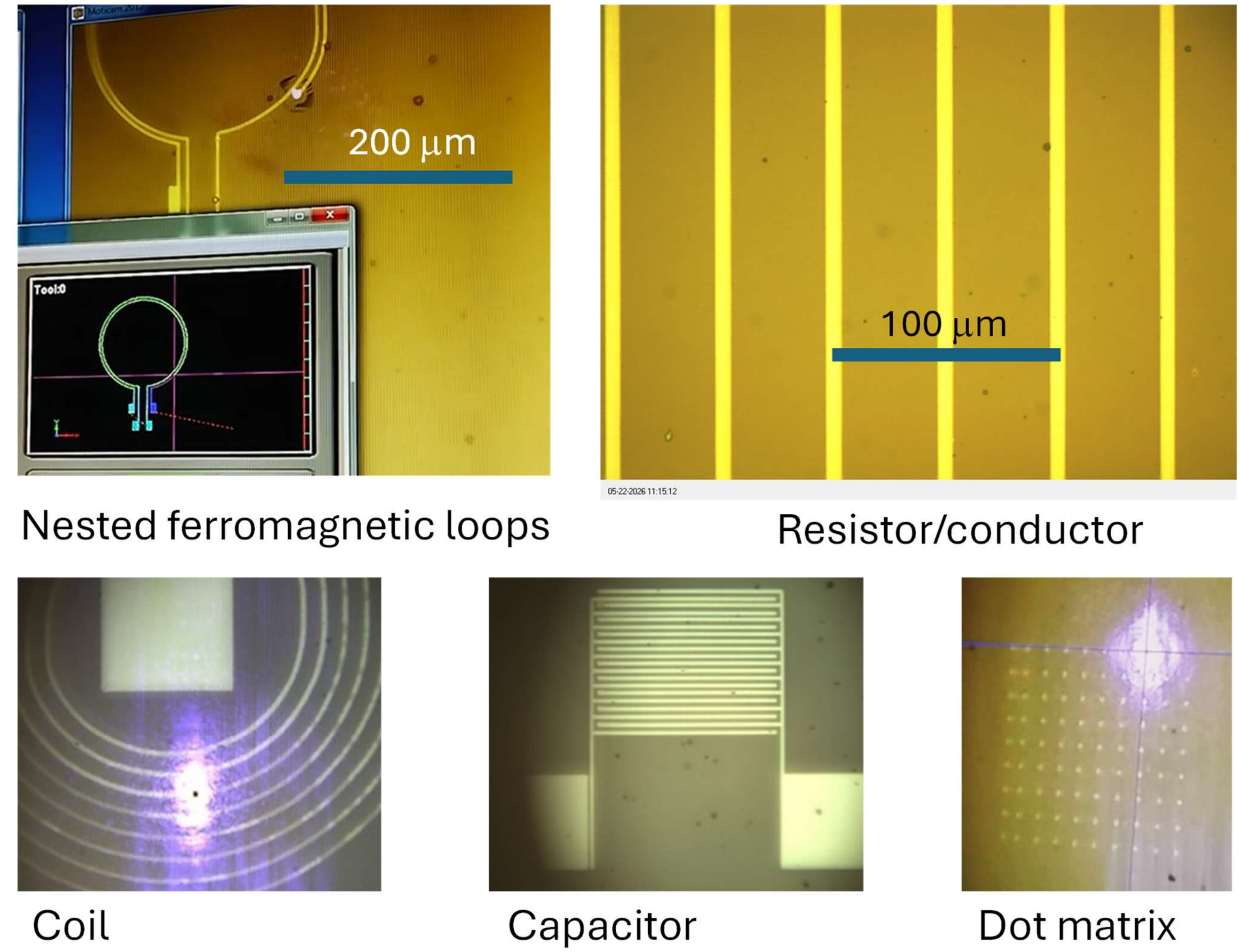

Figure 6. Examples of thin-film devices fabricated by direct redox patterning by 405nm laser, showing metallic features embedded within the insulating oxide matrix. The demonstrated structures include nested ferromagnetic loops, resistive or conductive lines, coils, capacitors, and dot-matrix arrays.